\documentclass[aps,prb,final,twocolumn,superscriptaddress,showkeys,showpacs]{revtex4}

\usepackage{graphicx}
\usepackage{dcolumn}
\usepackage{bm}
\usepackage{amsmath}
\usepackage{amssymb}

\newcommand\abbwidth{0.8\linewidth}

\newcommand\SA{B.1~}
\newcommand\SB{B.2~}
\newcommand\SC{A.1~}
\newcommand\SD{A.2~}
\newcommand\cm{\textrm{cm}^{-1}}

\usepackage{color}

\begin{document}
\title{Edge induced magneto plasmon excitation in a two-dimensional electron gas under quantum Hall conditions}

\author{Christian Notthoff}
\email[]{christian.notthoff@uni-due.de}
\affiliation{Fachbereich Physik and CeNIDE, Universit\"at Duisburg-Essen, Lotharstr.1, 47048 Duisburg, Germany}
\affiliation{Nanoparticle Process Technology, Universit\"at Duisburg-Essen, Lotharstr.1, 47048 Duisburg, Germany}
\author{Dirk Reuter}
\author{Andreas D. Wieck}
\affiliation{Lehrstuhl f\"ur Angewandte Festk\"orperphysik, Ruhr-Universit\"at Bochum, Universit\"atsstra{\ss}e 150, 44780 Bochum, Germany}
\author{Axel Lorke}
\affiliation{Fachbereich Physik and CeNIDE, Universit\"at Duisburg-Essen, Lotharstr.1, Duisburg 47048, Germany}

\date{\today}

\begin{abstract}
The spectrally resolved terahertz photoconductivity between two separately contacted edge-channels of a two-dimensional electron gas in the quantum Hall regime is investigated. We use a not-simply-connected sample geometry which is topologically equivalent to a Corbino disk. Due to the high sensitivity of our sample structure, a weak resonance situated on the high-energy side of the well known cyclotron resonance is revealed. The magnetic field as well as the carrier density dependence of this weak resonance, in comparison with different models suggests that the additional resonance is an edge induced magneto plasmon.
\end{abstract}


\preprint{in preparation for PRB}
\maketitle
\section{Introduction}
In recent years, two-dimensional electron gases (2DEGs) in the quantum Hall (QH) regime have attracted considerable interest in connection with the possibility of developing highly sensitive terahertz (THz) detectors based on the photo-induced change in conductivity\cite{PRB76,MJ36,EP2DS,APL87}. Unlike conventional absorption experiments, this technique combines magneto transport with spectroscopy and provides us with a unique tool to investigate not only the bulk cyclotron resonance (CR) of the 2DEG. It is now possible to resolve even spectral features, which are related to effects of the edge states in the QH effect\cite{PRB63,Lorke,IchPRB}, as well as excitations of collective magneto plasmon modes\cite{PRL93,PRB63MP,PRB48} and spin-flip transitions\cite{PRB67,EPL63}.\\
In the present paper, we use spectrally resolved photoconductivity measurements on a 2DEG formed in a GaAs/AlGaAs heterostructure to investigate the THz excitation process in the QH regime. We use a sample structure, which is topologically equivalent to a Corbino geometry, however, the circumference is greatly enlarged by a meander-like patterning. This sample structure provides us with a high sensitive of $\sim10^{7}$~V/W for THz radiation\cite{EP2DS} and allows us to investigate not only one dominant resonance -- which we identify as the bulk cyclotron resonance (CR) -- but reveals also a second, very weak resonance, situated on the high-energy side of the CR peak. We will present measurements on the magnetic field dispersion and the carrier density dependence of this high-energy resonance. Additionally, we will present measurements in tilted magnetic fields, which exclude spin effects as the origin of the high-energy resonance. Finally, we show that the high-energy resonance observed in the present experiments can be interpreted as a unusual magneto plasmon excitation, where the characteristic length scale of the plasmon modes is given by the depletion length of the 2DEG at the sample boundary.
\section{Experimental Setup}\label{sec:expsetup}
The samples are fabricated from two different molecular-beam-epitaxially grown GaAs/AlGaAs heterostructures A and B, using standard photolithography and wet chemical etching. The 2DEG of wafer A is located 60~nm below the surface. The carrier concentration is $N_e=4.6\times10^{11}cm^{-2}$ and the mobility of is approximately $\mu=500000~cm^2/Vs$. The 2DEG in wafer B is located 110~nm below the surface and has a carrier concentration and a mobility of approximately $N_e= 2.61\times10^{11}cm^{-2}$ and $\mu=288100~cm^2/Vs$, respectively. Ohmic contacts are provided by alloyed AuGe/Ni/AuGe~$(88:12)$ pads.
\begin{figure}[htb]
  \begin{center}\leavevmode
    \includegraphics[width=\abbwidth]{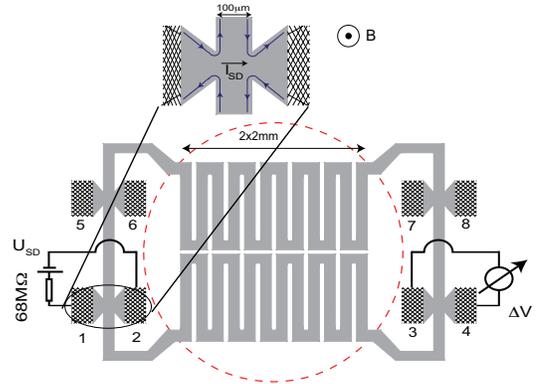}
    \caption{(Color online) Sketch of the sample geometry. Contacts (hatched) are positioned along the inner and outer edges of the meander-like patterned mesa (gray). Arrows indicate the direction of electron drift motion in the edge-channels for an integer filling factor.}\label{fig:sample_setup}
  \end{center}
\end{figure}\\
Figure \ref{fig:sample_setup} schematically shows the topology of the samples, consisting of a meander-like patterned (Corbino-geometry) mesa, with a circumference (total length of the edges) of $\sim32$ mm for all samples. The channel width of samples \SC, \SD, and \SA is $w=100~\mu$m, whereas sample \SB has a width of $w=50~\mu$m. Sample \SD is furthermore equipped with a NiCr top gate to tune the carrier density. The large circle in Fig. \ref{fig:sample_setup} indicates the illuminated area. \\
In the present experiment, we record the photo-induced change in conductivity by applying a constant current between the inner and outer edge (contacts 1 and 2 in Fig.~\ref{fig:sample_setup}) and measuring the change of the non-local voltage $\Delta V$ between two other, well-separated contacts (3 and 4). We use a Bruker IFS113v Fourier-transform spectrometer (FTS), and the sample itself as a photo detector to obtain spectrally resolved photoresponse measurements. At fixed magnetic fields, the broadband THz radiation of a mercury lamp was modulated by the FTS and the corresponding change in the non-local voltage $\Delta V$ was ac coupled to a high-impedance ($\geq 100~\text{M}\Omega$) voltage preamplifier. The preamplified signal was recorded as an interferogram, which was Fourier transformed to obtain the photoconductivity spectrum. In the spectral range of interest, the light source and all optical components have a very weak dependence on wavelength, so that the recorded signal will be dominated by the spectral response of the sample itself. A system of mirrors and polished stainless-steel tubes, acting as oversized waveguides, guides the THz radiation to the sample, mounted inside a $^4$He cryostat with a superconducting magnet, to apply fields up to $B=12$~T. Optionally, a commercial Si bolometer can be mounted below the sample to perform transmission measurements. All data were obtained by $T=4.2$~K in the Faraday geometry.
\section{Experimental results}\label{sec:expres}
\begin{figure}[htb]
  \begin{center}\leavevmode
    \includegraphics[width=\abbwidth]{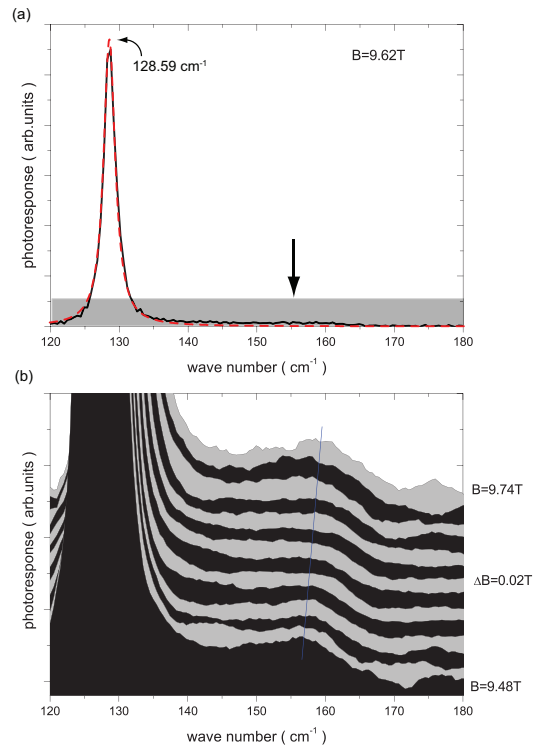}
    \caption{(Color online) (a) Photoconductivity spectrum of sample \SC measured at a magnetic field of $B=9.62$~T (solid line) and a Lorentzian fit to the data (dashed red line). (b) Normalized photoconductivity spectra between $B=9.48$~and $9.74$~T, where the scaling corresponds to the gray shaded area in (a). The spectra are shifted along the $y$ axis for clarity and the line is a guide to the eye.}\label{fig:b_sweep_schulter}
  \end{center}
\end{figure}

Figure \ref{fig:b_sweep_schulter} (a) shows a typical THz photoconductivity spectrum of sample \SC at a magnetic field of $B=9.62$~T (filling factor $\nu=2$). A pronounced resonance is observed, which can be fitted by a single Lorentzian profile (dashed red line). The fit yields a resonance frequency of $128.59~\cm$ and a full width at half maximum of $2~\cm$.
\\
A closer look at the data reveals a second, very weak resonance at the high-energy side of the dominant peak, indicated by an arrow in Fig.~\ref{fig:b_sweep_schulter} (a). For a better analysis, we plot in Fig.~\ref{fig:b_sweep_schulter} (b) the enlarged spectra measured between $B=9.48$~and $9.74$~T. All the spectra are normalized to their dominant peak to account for the photoconductivity sensitivities at different magnetic fields so that we can directly compare the intensity of the second resonance. As indicated by the line in Fig.~\ref{fig:b_sweep_schulter} (b), the resonance position shifts to higher energies with increasing magnetic field whereas the amplitude remains constant.
\begin{figure}[htb]
  \begin{center}\leavevmode
    \includegraphics[width=\abbwidth]{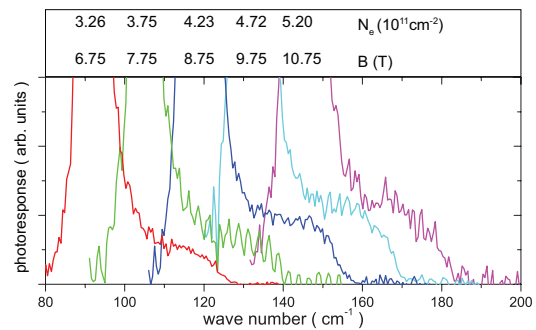}
    \caption{(Color online) Normalized photoconductivity spectra of sample \SD at different carrier densities, while the magnetic field is adjusted to keep the filling factor $\nu=2$. }\label{fig:B_sweep_N_sweep_CBR_dispersion}
  \end{center}
\end{figure}\\
Figure~\ref{fig:B_sweep_N_sweep_CBR_dispersion} shows typical THz photoconductivity spectra of sample \SD at different carrier densities, while the magnetic field is adjusted to keep the filling factor $\nu=2$ and maximize the photoconductivity sensitivities\cite{EP2DS}. Again all spectra are normalized to their dominant peak. Here we observe a slight narrowing of the dominant peak with decreasing carrier density, which causes the amplitude of the high-energy resonance to decrease apparently. But a closer analysis (taking into account the oscillator strength of the main peak) shows that the strength of the high-energy resonance is independent of the carrier density as well as of the magnetic field.
\begin{figure}[htb]
  \begin{center}\leavevmode
    \includegraphics[width=\abbwidth]{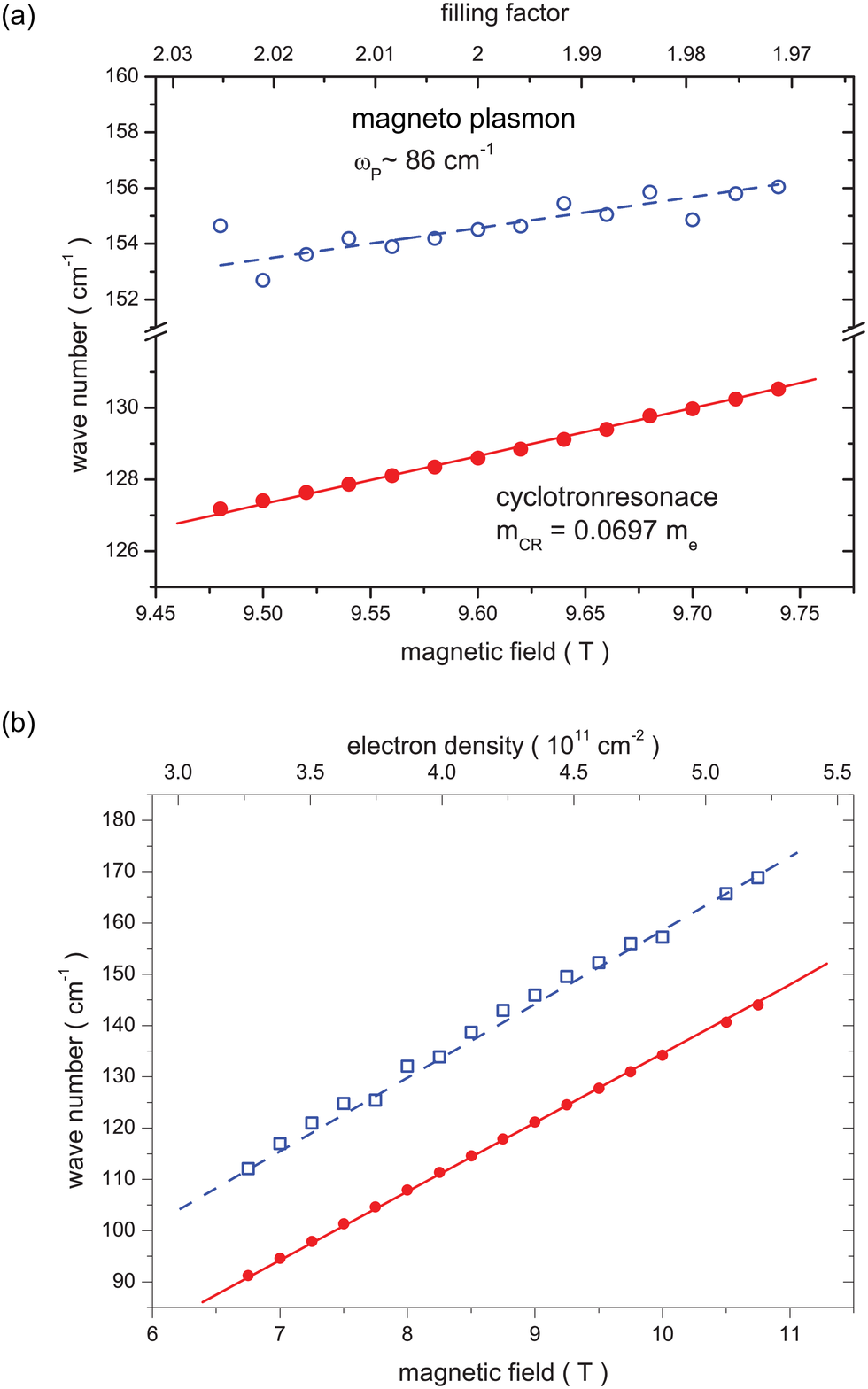}
    \caption{(Color online) (a) Magnetic field dispersion of the CR (solid red circles) and the high-energy resonance (open blue circles) obtain from fitting the photoconductivity spectra shown in Fig.~\ref{fig:b_sweep_schulter}. The dashed blue line shows the fit of Eq. (\ref{eq:def_mp}) using a plasmon frequency of $\omega_p=86~\cm$. (b) Carrier density dependence of the CR (solid red circles) and the high-energy resonance (open blue squares) obtain by fitting the spectra shown in Fig. \ref{fig:B_sweep_N_sweep_CBR_dispersion}. The dashed line is a guide to the eye. The solid red lines in (a) and (b) correspond to $\omega_{c}=eB/m^*$ using an effective mass of $m^*=0.0697~m_e$ .}\label{fig:B_dispersion_MP}
  \end{center}
\end{figure}\\
In Figs.~\ref{fig:B_dispersion_MP}(a) and \ref{fig:B_dispersion_MP}(b) we plot the dispersion of both resonances determined from the photoconductivity spectra shown in Figs.~\ref{fig:b_sweep_schulter} and \ref{fig:B_sweep_N_sweep_CBR_dispersion}, respectively. The magnetic field dispersion of the dominant peak (solid red circles) corresponds to the bulk cyclotron resonance (CR) dispersion (solid red lines), with $\omega_{c}=eB/m^*$ using an effective mass of $m^*=0.0697~m_e$ (see, e.g., Refs. \onlinecite{STC66,PR122,PRB53,bnp}). As expected for the CR, the influence of the carrier concentration on the effective mass is negligible [Fig. \ref{fig:B_dispersion_MP} (b)]. Furthermore, the resonance width matches exactly the width deduced from the transport mobility and the carrier density of sample \SC (see, e.g., Ref.~\onlinecite{SURF58}). Therefore, the dominant resonance can be easily identified as a bolometric change in conductivity due to a heating of the electron gas by the bulk cyclotron absorption.\\
The dispersion of the high-energy resonance also appears to be linear in Fig. \ref{fig:B_dispersion_MP}. Below, we will discuss that its behavior is more complicated. But in contrast to the CR, the high-energy resonance shows a reduced slope in the magnetic field dispersion [Fig. \ref{fig:B_dispersion_MP} (a)] while a slightly increased slope is observed for the carrier-density-dependent measurements [open blue squares in Fig. \ref{fig:B_dispersion_MP} (b)]. The carrier density dependence of the high-energy resonance indicates a many-body effect.
\begin{figure}[htb]
  \begin{center}\leavevmode
    \includegraphics[width=\abbwidth]{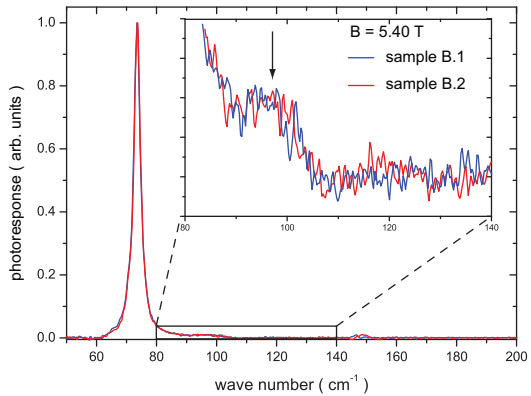}
    \caption{(Color online) Photoconductivity spectra of samples \SA (blue curve) and \SB (red curve) measured at a magnetic field of $B=5.40$~T. The mesa structure of sample \SA has a channel width of $100~\mu$m while sample \SB is patterned with a $50~\mu$m channel width. The spectra are normalized to the dominant CR resonance amplitude.}\label{fig:vgl_l_m}
  \end{center}
\end{figure}\\
To investigate a possible influence of the geometrical confinement of the 2DEG on the high-energy resonance, we have performed spectrally resolved  photoconductivity measurements on samples with two different channel widths. Figure \ref{fig:vgl_l_m} shows a typical spectrum of sample \SA with a channel width of $w=100~\mu$m (blue curve) in comparison to a spectrum of sample \SB at the same magnetic field while the channel width is reduced to $w=50~\mu$m (red curve). Neither the position of the dominant CR resonance nor the high-energy resonance (indicated by an arrow) are influenced by the lateral, geometrical confinement.\\
In the following section we are going to discuss two different approaches to explain the occurrence of the high-energy resonance observed in the present experiments.
\section{Discussion}
Figures~\ref{fig:b_sweep_schulter} (b) and~\ref{fig:B_sweep_N_sweep_CBR_dispersion} show a remarkable similarity with the photoconductivity spectra obtained from InAs quantum wells reported by [see Fig.~3 (b) in Ref.~\onlinecite{PRB67}]. They identify the high-energy resonance as a combined resonance (CBR). Due to the similarity between the spectra observed in Ref.~\onlinecite{PRB67} and our present spectra, we try to fit the dispersion of the high-energy resonance (open blue circles in Fig.~\ref{fig:B_dispersion_MP}) also in terms of a CBR. The resonance frequency of the CBR is given by \cite{PRB41}
\begin{equation}\label{eq:def_cbr}
\omega_{CBR}=\sqrt{(\omega_c+\omega_z)^2+(2\alpha k_F/\hbar)^2}~,
\end{equation}
with the CR frequency $\omega_c=eB/m^*$, the electron-spin-resonance frequency $\omega_z=-g~e~B/(2m_e)$, the spin-orbit parameter $\alpha$ and the Fermi wave vector $k_F=\sqrt{2\pi N_e}$. The dispersion of the high-energy resonance (Fig. \ref{fig:B_dispersion_MP}) can be fitted in different ways using Eq.~(\ref{eq:def_cbr}). On the one hand, the spin-orbit parameter $\alpha$ can be used as a free parameter, taking the well-known $g$~factor of $-0.44$ for bulk GaAs\cite{PRL51,PRB15}. Alternatively, we can take both $\alpha$ and $g$ as fitting parameters. In both cases, the fitting procedure yields a spin-orbit parameter of approximately $\alpha=3\times 10^{-11}$~eVm, which is an order larger then the theoretically predicted one for bulk GaAs ($\alpha\approx10^{-13}$~eVm)\cite{PRB51}. On the other hand, the data can also be fitted assuming a reasonable spin-orbit parameter of $\sim10^{-13}$~eVm and a strongly enhanced $g$~factor of $g=-6$.
\begin{figure}[htb]
  \begin{center}\leavevmode
    \includegraphics[width=\abbwidth]{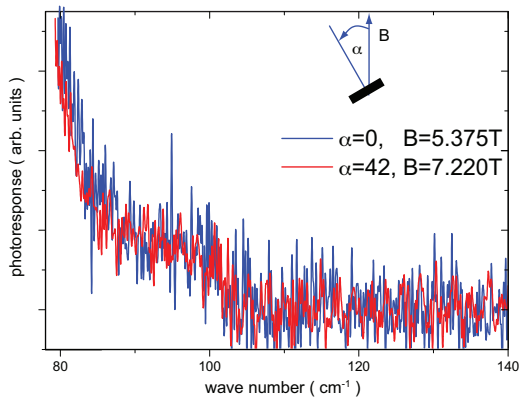}
    \caption{(Color online) Photoconductivity spectrum of sample \SA measured at a magnetic field of $B=5.38$~T (blue curve) and the field direction parallel to the surface normal, comparing a spectrum measured at $B=7.22$~T and a tilt angle of $42^{\circ}$ (red curve).}\label{fig:tilted_field}
  \end{center}
\end{figure}\\
To rule out the possibility of such an enhanced $g$~factor (e.g., observed by Nicholas $et~al$.\cite{PRB37}) as the origin of the high-energy resonance we use the fact that, in a tilted magnetic field, the cyclotron resonance position only depends on the magnetic field component normal to the 2DEG while a spin contribution should depend on the total magnetic field\cite{PRB37}. In Fig.~\ref{fig:tilted_field} we compare two photoconductivity spectra of sample \SA,  one with the magnetic field direction normal to the surface (blue curve) and the other with a magnetic field tilted by $\sim42^{\circ}$ (red curve).
For a direct comparison the magnetic fields are chosen, so that the parallel magnetic field components of both measurements and therefore the CR positions are identical. Using Eq. (\ref{eq:def_cbr}) with a spin-orbit parameter  of $\alpha\approx10^{-13}$~eVm~(Ref. \onlinecite{PRB51}) and a $g$~factor resulting from the above fit of the magnetic field dispersion ($g=-6$) we would expect at least a shift of $\Delta \omega=5~\cm$ to higher frequencies for the spectrum taken in the tilted magnetic field. Experimentally, however, we do not observe any shift of the high-energy resonance for the tilted magnetic field measurement (see Fig. \ref{fig:tilted_field}). This clearly rules out an enhanced $g$~factor as the origin of the high-energy resonance observed in the present experiment. Despite the similarity of our photoconductivity spectra [Figs. \ref{fig:b_sweep_schulter} (b) and \ref{fig:B_sweep_N_sweep_CBR_dispersion}] with the spectra obtain from InAs quantum wells\cite{PRB67}, the assumption of a CBR could not explain the high-energy resonance observed in the present experiment on GaAs heterostructures.\\
Therefore, a different approach is needed to explain the observations in the present work. In the following, we suggest to interpret the high-energy resonance in terms of a collective many-body excitation, i.e. the magneto plasmon resonance (MP). The $B$-field dispersion of the MP is given by \cite{PRB48,PRL90,PRB34}
\begin{equation}\label{eq:def_mp}
\omega_{MP}=\sqrt{\omega_c^2+\omega_p^2}~,
\end{equation}
with the plasmon frequency at zero magnetic field\cite{PRL18}
\begin{equation}\label{eq:def_p}
\omega_{p}=\sqrt{\frac{N_ee^2}{2m^*\varepsilon_0\varepsilon_{eff}}k}~,
\end{equation}
where $k$ is the plasmon wave vector, $N_e$ is the 2D electron density, and $\varepsilon_{eff}$ is the effective dielectric constant. Performing photoconductivity measurements on conventional Hall bar samples allows for the investigation of the MP without the need of a grating to couple the incident radiation with plasmons of a wave vector $k$. As described, e.g., by Vasiliadou $et~al$.,\cite{PRB48} the boundaries of a Hall bar themselves provides the coupling and determine the frequency of the confined plasmon. Their measurements are in a good agreement with calculations using the Eqs.~(\ref{eq:def_mp}) and (\ref{eq:def_p}) with $k=\pi/w$, where $w$ is the width of the Hall bar. But as shown in the previous section (see Fig. \ref{fig:vgl_l_m}), no shift of the high-energy resonance can be observed in the present experiment if the channel width is reduced from $w=100~\mu$m (sample \SA) to $w=50~\mu$m (sample \SB), while a shift by the factor of $\sqrt{2}$ is expected for $\omega_p$~[see Eq. \ref{eq:def_p}]. Furthermore, we would expect to find the confined plasmon frequency in the GHz regime for our samples\cite{PRB48}. However, fitting the magnetic field dispersion of the high-energy resonance [dashed blue line in Fig. \ref{fig:B_dispersion_MP} (a)], using Eq.~(\ref{eq:def_mp}), yields a plasmon frequency of $\omega_p=86~\cm$, corresponding to $2.6$~THz. For the MP picture to hold we need a characteristic length scale which is independent of the geometrical definition of the mesa and which is short enough to cause a plasmon frequency in the THz regime.\\
As reported by Elliott $et~al$.\cite{PRB56} an edge-induced magneto plasmon resonance (EIMP) can be observed in measurements on a circular sheet of $^4$He$^+$ ions, where the resonance frequency is determined by the depletion length of the electron density profile at the sample edge.\\
The depletion length of the carrier density of our GaAs/AlGaAs heterostructure is given by\cite{CSG}
\begin{equation} \label{eq:def_lcsg}
l_{CSG}=\frac{E_g\varepsilon_{GaAs}\varepsilon_0}{\pi e^2N_e}~,
\end{equation}
where $E_g=1.52$~eV is the energy gap of GaAs at a temperature of 4.2 K. Using the depletion length $l_{CSG}$ and substituting $k=\frac{3}{4}\pi / l_{CSG}\approx 2.4/l_{CSG}$~(Ref.~\onlinecite{PRB34}) in Eq.~(\ref{eq:def_p}) we obtain a EIMP frequency of $\omega_{EIMP}=85~\cm$ for sample \SC, which is  in a good agreement with the experimentally determined resonance frequency $\omega_p=86~\cm$. Here we have used $\varepsilon_{eff}=\frac{1}{2}\varepsilon_{GaAs}(1+\coth{(kh)})$ (Refs. \onlinecite{Chaplik} and \onlinecite{Surf170}) for the effective dielectric constant, where $\varepsilon_{GaAs}=12.8$ is the dielectric constant of GaAs\cite{PRL90,PRB34,PRB48} and $h$ is the depth of the 2DEG below the sample surface.
\begin{figure}[htb]
  \begin{center}\leavevmode
    \includegraphics[width=\abbwidth]{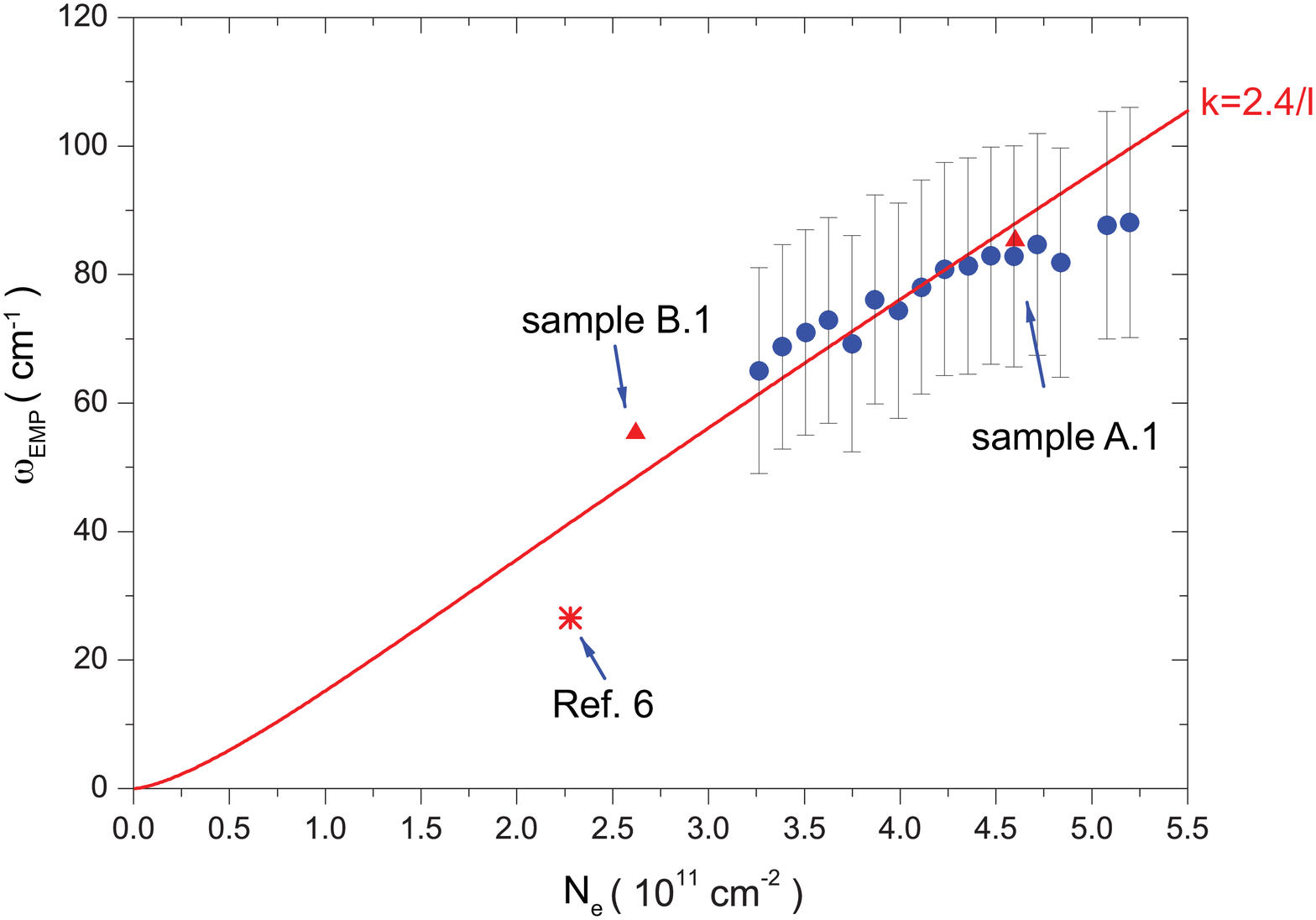}
    \caption{(Color online) Carrier density dependence of the EIMP resonance frequency obtained from fitting the photoconductivity spectra shown in  Fig.~\ref{fig:B_sweep_N_sweep_CBR_dispersion} (points) and the theoretical predicted dispersion according to Eqs. (\ref{eq:def_p}) and (\ref{eq:def_lcsg}) (line).}\label{fig:B_sweep_N_sweep_MP_dispersion}
  \end{center}
\end{figure}\\
Figure \ref{fig:B_sweep_N_sweep_MP_dispersion} shows the experimentally determined resonance frequency as a function of the carrier density $N_e$, evaluated from the data of sample \SD shown in Fig. \ref{fig:B_dispersion_MP} (b) using Eq. (\ref{eq:def_mp}) (solid blue circles). Additionally, the resonance frequency obtained from samples \SC and \SA (red triangles) as well as the plasmon frequency determined by a photocurrent experiments described in Ref. \onlinecite{Lorke} (red star) are plotted. The solid red line represents the calculated EIMP frequency as a function of the carrier density using Eqs.  (\ref{eq:def_p}) and (\ref{eq:def_lcsg}). The calculated values resembles the experimentally determined ones quite well, without the need for any fit parameter. The agreement over a broad range of carrier densities and also for different samples suggests that the origin of the high-energy resonance observed in the present experiment is indeed the excitation of a collective magneto plasmon oscillation, where the plasmon wave vector and thus the resonance frequency $\omega_p$ is given by the depletion length at the sample edge. So far, the microscopic excitation mechanism at the sample boundary and the influence of the compressible and incompressible stripes at different filling factors are not fully understood. As shown, e.g., in Ref. \onlinecite{IchPRB}, the edge reconstruction in the quantum Hall regime can have a pronounced influence on the photoconductive properties of 2DEGs beyond the collective bulk excitations. Further theoretical work on THz response of the reconstructed edge in the quantum Hall regime is therefore highly desirable.
\section{Conclusion}
In conclusion, we have investigated the weak THz photoresponse, situated on the high-energy side of the cyclotron resonance, using spectrally resolved photoconductivity measurements in the quantum Hall regime. We find that the magnetic field and the carrier density dependence of the resonance position is in a good agreement with an edge-induced magneto plasmon excitation, where the characteristic length scale (plasmon wave vector $k$) is determined by the depletion length of the 2DEG at the sample boundary.
\begin{acknowledgments}
We wish to acknowledge financial support by the Deutsche Forschungsgemeinschaft (DFG, projects No. LO 705/1-3 and No. SFB 508).
\end{acknowledgments}


\begin{thebibliography}{9}
\bibitem{PRB76} K. Ikushima, H. Sakuma, S. Komiyama, and K. Hirakawa, {Phys. Rev. B} \textbf{76}, 165323 (2007).
\bibitem{MJ36} Y. Kawano, and T. Okamoto, {Microelectronics Journal} \textbf{36}, 592 (2005).
\bibitem{EP2DS} C. Notthoff, A. Lorke, and D. Reuter, {Physica E} \textbf{40}, 1328 (2008).
\bibitem{APL87} C. Stellmach, A. Hirsch, G. Nachtwei, Yu. B. Vasilyev, N. G. Kalugin, and G. Hein, {App. Phys. Lett.} \textbf{87}, 133504 (2005).
\bibitem{PRB63} K. Hirakawa, K. Yamanaka, Y. Kawaguchi, M. Endo, M. Saeki, and S. Komiyama, {Phys. Rev. B} \textbf{63}, 085320 (2001).
\bibitem{Lorke} A. Lorke, J. P. Kotthaus, J. H. English, and A. C. Gossard, {Phys. Rev. B} \textbf{53}, 1054 (1996).
\bibitem{IchPRB} C. Notthoff, K. Rachor, D. Heitmann, A. Lorke, {Phys. Rev. B} \textbf{80}, 205320 (2009).
\bibitem{PRL93} S. Holland, Ch. Heyn, D. Heitmann, E. Batke, R. Hey, K. J. Friedland, and C. -M. Hu, {Phys. Rev. Lett.} \textbf{93}, 186804 (2004).
\bibitem{PRB63MP} B. G. L. Jager, S. Wimmer, A. Lorke, J. P. Kotthaus, W. Wegscheider, and M. Bichler, Phys. Rev. B \textbf{63}, 045315 (2001).
\bibitem{PRB48} E. Vasiliadou, G. M\"uller, D. Heitmann, D. Weiss, K. v. Klitzing, H. Nickel, W. Schlapp, and R. L\"osch, {Phys. Rev. B} \textbf{48}, 17145 (1993).
\bibitem{PRB67} C. -M. Hu, C. Zehnder, Ch. Heyn, and D. Heitmann, {Phys. Rev. B} \textbf{67}, 201302 (2003).
\bibitem{EPL63} C. Zehnder, A. Wirthmann, Ch. Heyn, D. Heitmann, and C.-M. Hu, {Europhys. Lett.} \textbf{63}, 576 (2003).
\bibitem{STC66} E.B. Hansen and O.P. Hansen, {Solid State Comm.} \textbf{66}, 1181 (1988).
\bibitem{PR122} E. D. Palik, G. S. Picus, S. Teitler, and R. F. Wallis, {Phys. Rev.} \textbf{122}, 475 (1961).
\bibitem{PRB53} P. Pfeffer and W. Zawadzki, {Phys. Rev. B} \textbf{53}, 12813 (1996).
\bibitem{bnp} F. Thiele, U. Merkt,  J. P. Kotthaus, G. Lommer, F. Malcher, U. R\"ossler, and G. Weimann, {Solid State Commun.} \textbf{63}, 841 (1987).
\bibitem{SURF58} K.W. Chiu, T.K. Lee, and J.J. Quinn, Surface Science \textbf{58}, 182 (1976).
\bibitem{PRB41} B. Das, S. Datta, and R. Reifenberger, {Phys. Rev. B} \textbf{41}, 8278 (1990).
\bibitem{PRL51} D. Stein, K. v. Klitzing, and G. Weimann, {Phys. Rev. Lett.} \textbf{51}, 130 (1983).
\bibitem{PRB15} C. Weisbuch, and C. Hermann, {Phys. Rev. B} \textbf{15}, 816 (1977).
\bibitem{PRB51} F.G. Pikus, and G. E. Pikus, {Phys. Rev. B} \textbf{51}, 16928 (1995).
\bibitem{PRB37} R. J. Nicholas, R. J. Haug, K. Klitzing, and G. Weimann,{Phys. Rev. B} \textbf{37}, 1294 (1988).
\bibitem{PRL90} I. V. Kukushkin, J. H. Smet, S. A. Mikhailov, D. V. Kulakovskii, K. v. Klitzing, and W. Wegscheider, {Phys. Rev. Lett.} \textbf{90}, 156801 (2003).
\bibitem{PRB34} Leavitt, P. Richard, and J.W. Little, {Phys. Rev. B} \textbf{34}, 2450 (1986).
\bibitem{PRL18} F. Stern, {Phys. Rev. Lett.} \textbf{18}, 546 (1967).
\bibitem{PRB56} P.L. Elliott, S.S. Nazin, C.I. Pakes, L. Skrbek, W.F. Vinen, and G.F. Cox, {Phys. Rev. B} \textbf{56}, 3447 (1997).
\bibitem{CSG} D. B. Chklovskii, B. I. Shklovskii, and L. I. Glazman, {Phys. Rev. B} \textbf{46}, 4026 (1992).
\bibitem{Chaplik} A.V. Chaplik, {Surface Science Reports} \textbf{5}, 289 (1985).
\bibitem{Surf170} D. Heitmann, {Surface Science} \textbf{170}, 332 (1986).
\end{thebibliography}
\end{document}